\documentclass[preprint,prd,noshowpacs]{revtex4}
\usepackage{amssymb}
\usepackage{amsmath}
\usepackage{amsfonts}
\usepackage{graphicx, float}

\begin{document}

\title{Scalar gauge fields}

\author{Eduardo I. Guendelman}
\email{guendel@bgu.ac.il}
\affiliation{Physics Department, Ben Gurion University of the Negev, Beer Sheva 84105, Israel \\
and \\
Department of Physics, California State University, Fresno, CA 93740-8031 USA}
\author{Douglas Singleton}
\email{dougs@csufresno.edu}
\affiliation{Department of Physics, California State University, Fresno, CA 93740-8031 USA}

\date{\today}

\begin{abstract}
In this paper we give a variation of the gauge procedure which employs a {\it scalar} gauge field, $B (x)$, 
in addition to the usual {\it vector} gauge field, $A_\mu (x)$. We study this variant of the usual 
gauge procedure in the context of a complex scalar, matter field  $\phi (x)$ with a $U(1)$ symmetry. 
We will focus most on the case when $\phi$ develops a vacuum expectation value via spontaneous symmetry
breaking. We find that under these conditions the 
scalar gauge field mixes with the Goldstone boson that arises from the breaking of a global symmetry. 
Some other interesting features of this scalar gauge model are: (i) The new gauge 
procedure gives rise to terms which violate $C$ and $CP$ symmetries. This may have 
have applications in cosmology or for CP violation in particle physics; (ii) the existence of 
mass terms in the Lagrangian which respect the new extended gauge symmetry. Thus one can have gauge field mass 
terms even in the absence of the usual Higgs mechanism; (iii) the emergence of a sine-Gordon
potential for the scalar gauge field; (iv) a natural, axion-like suppression of the interaction strength of the
scalar gauge boson.
\end{abstract}

\maketitle

\section{Introduction}

In this work we give an extension of the usual idea of gauging (i.e. making local) a symmetry. 
We lay out the basic structure of the model without having some specific application in mind. The justification 
for looking into variations of the standard gauge process, without having a specific application in
mind, is the centrality of the standard gauge procedure to all of modern physics. We will show
that when spontaneous symmetry breaking occurs the theory exhibits a physical particle/field which is
a mixture of the scalar gauge field and the Goldstone boson. In laying out the
general frame work of this model we will mention some interesting features of this variant of the gauge 
procedure: (i) It gives a novel mechanism for $C$ and  $CP$ violation; (ii) it has mass generation for the vector gauge 
boson without having to use the standard Higgs mechanism; (iii) one can naturally obtain a sine-Gordon
equation of motion for the scalar field with its associated soliton/kink solutions; (iv) the interaction 
of the scalar gauge boson/Goldstone boson can be made weak via a natural, axion-like mechanism. 
We discuss the physical reality of this new scalar gauge bosons -- whether it is a physically observable field or
if it is some auxiliary, non-dynamical field. We show that this scalar gauge field is closely related to the
Goldstone boson and in the case when spontaneous symmetry break occurs that the one finds a physical particle which
is a mixture of the Goldstone boson and the scalar gauge field.

We begin by recalling the usual gauge procedure which starts with a scalar, matter field, $\phi$, or spinor field,
$\Psi$, which has some global phase symmetry. Then by making this global phase symmetry a local symmetry
one finds that one needs to introduce a vector gauge field, $A_\mu$. Although our 
alternative gauge procedure works for spinor fields and for non-Abelian gauge symmetries, for simplicity we 
will consider a complex scalar field, $\phi$, with a $U(1)$ symmetry which has the Lagrange density 
\begin{equation}  \label{scalar-L}
{\cal L} = \partial _\mu \phi \partial ^\mu \phi ^* - V(\phi)~,
\end{equation}
where $V(\phi )$ is some scalar self interaction potential. A typical example is 
to have a Higgs, symmetry breaking potential of the form 
$V(\phi) = -m^2 |\phi| ^2 + \lambda |\phi |^4$ where $|\phi |^2 = \phi^* \phi$.
Now the Lagrangian density in \eqref{scalar-L} has the global gauge symmetry
$\phi \rightarrow e^{i e \Lambda} \phi$ where $e$ is the electric charge of the scalar field
and $\Lambda$ is a global phase parameter. If one lets the phase parameter become space-time
dependent (i.e. $\Lambda \rightarrow \Lambda (x_\mu )$) one can still maintain this new local 
phase symmetry by introducing a four-vector gauge field $A_\mu$. First one replaces the ordinary derivative by
covariant derivatives, $D_\mu$, of the form
$$
\partial _\mu \rightarrow D_\mu = \partial _\mu + i e A_\mu ~,
$$ 
along with the requirement that, in addition to the phase transformation of $\phi$ 
(i.e. $\phi \rightarrow e^{i e \Lambda (x_\mu )} \phi$), the four-vector $A_\mu$ transform
like $A_\mu \rightarrow A_\mu - \partial _\mu \Lambda$. One can see that 
$D_\mu \phi$ , $(D_\mu \phi )^*$ are covariant under these two transformations which together
form the $U(1)$ gauge transformation. Additionally one can add a term to \eqref{scalar-L}
which involves only $A_\mu$ and is invariant under the gauge transformation. The Maxwell field
strength tensor is such a quantity
\begin{equation}
\label{maxwell-ft}
F_{\mu \nu} = \partial _\mu A_\nu - \partial _\nu A_ \mu ~.
\end{equation} 
One can see that $F_{\mu \nu}$ is invariant under $A_\mu \rightarrow A_\mu - \partial _\mu \Lambda$.
Putting all of the above together leads to the following Lagrangian
\begin{equation}
\label{u1}
{\cal L} = D _\mu \phi (D ^\mu \phi )^* - V(\phi) - \frac{1}{4} F_{\mu \nu} F^{\mu \nu}~,
\end{equation}
with $D_\mu$ and $F_{\mu \nu}$ defined above. This Lagrangian represents a complex, charged scalar
field $\phi$ coupled to a vector gauge boson, $A_\mu$. It respects the local gauge transformation
$$
\phi \rightarrow e^{i e \Lambda} \phi ~~~;~~~ A_\mu \rightarrow A_\mu - \partial _\mu \Lambda ~.
$$

\section{Scalar gauge field}

Starting with the same complex scalar field from \eqref{scalar-L} we now gauge the phase symmetry of
$\phi$ by introducing a real, scalar $B( x _\mu)$ and two types of covariant derivatives as
\begin{equation}
\label{cov-ab}
D ^A _\mu = \partial _\mu + i e A_\mu ~~~;~~~ D ^B _\mu = \partial _\mu + i e \partial _\mu B ~.  
\end{equation}  
The gauge transformation of the complex scalar, vector gauge field and scalar gauge field have the 
following gauge transformation
\begin{equation}
\label{gauge-trans}
\phi \rightarrow e^{i e \Lambda} \phi ~~~;~~~ A_\mu \rightarrow A_\mu - \partial _\mu \Lambda ~~~;~~~
B \rightarrow B - \Lambda ~.
\end{equation} 
It is easy to see that terms like $D ^A _\mu \phi$ and $D ^B _\mu \phi$, as well as their complex conjugates will be
covariant under \eqref{gauge-trans}. Thus one can generate kinetic energy type terms like 
$(D ^A _\mu \phi) (D ^{A \mu} \phi)^*$, $(D ^B _\mu \phi) (D ^{B \mu} \phi)^*$, $(D ^A _\mu \phi) (D ^{B \mu} \phi)^*$,
and $(D ^B _\mu \phi) (D ^{A \mu} \phi)^*$. Unlike $A_\mu$ where one can add a gauge invariant 
kinetic term involving only $A_\mu$ (i.e. $F_{\mu \nu} F^{\mu \nu}$) this is apparently not possible to do for the
scalar gauge field $B$. However note that the term $A_\mu - \partial _\mu B$ is invariant under the
gauge field transformation alone (i.e. $A_\mu \rightarrow A_\mu - \partial _\mu \Lambda$ and 
$B \rightarrow B - \Lambda$). Thus one can add a term like $(A_\mu - \partial _\mu B)(A^\mu - \partial ^\mu B)$
to the Lagrangian which is invariant with respect to the gauge field part only of the gauge transformation
in \eqref{gauge-trans}. This gauge invariant term will lead to both mass-like terms for the vector gauge
field and kinetic energy-like terms for the scalar gauge field. In total a general Lagrangian which respects 
the new gauge transformation and is a generalization of the usual gauge Lagrangian of \eqref{u1}, has 
the form
\begin{eqnarray}
\label{u2}
{\cal L} &=& c_1 D^A _\mu \phi (D ^{A \mu} \phi ) ^* + c_2 D^B _\mu \phi (D ^{B \mu} \phi )^* 
+ c_3 D^A _\mu \phi (D ^{B \mu} \phi )^* + c_4 D^B _\mu \phi (D ^{A \mu} \phi )^* - V(\phi) \nonumber \\ 
&-& \frac{1}{4} F_{\mu \nu} F^{\mu \nu} + c_5 (A_\mu - \partial _\mu B)(A^\mu - \partial ^\mu B)~,
\end{eqnarray}
where $c_i$'s are constants that should be fixed to get a physically acceptable Lagrangian. The arbitrariness
of this Lagrangian through the (at this point) undetermined $c_i$'s is inelegant, but at the beginning
we want to write down the most general Lagrangian which was consistent with the gauge transformation 
in \eqref{gauge-trans}. Later in section \eqref{B-potential} we will show that it is possible 
to add non-derivative , polynomial interaction terms for the $B$ field. As we progress through 
the paper we will try to give restrictions on the $c_i$'s which will reduce this arbitrariness.  
An interesting point to remark upon is that the last term in \eqref{u2}
will have a term of the form $A_\mu A^\mu$ which is a mass term for the vector gauge field. This term (contrary
to the usual gauge transformation given in the introduction) does not violate the expanded
gauge symmetry of \eqref{gauge-trans}. Also the last term in \eqref{u2} has a term of the form 
$\partial _\mu B \partial ^\mu B$ which is the usual kinetic energy term for a scalar field. Thus the
gauge field $B$ appears to be a dynamical field, at least at first glance. We will discuss
to what degree $B$ is a physical field later. The mass term for $A_\mu$ and a 
kinetic energy term for $B$ still respect the gauge transformation in \eqref{gauge-trans}.

Notice that although the local gauge symmetry is $U(1)$, the global symmetry is $U(1) \times U(1)$, since
the Lagrangian \eqref{u2} is invariant under the global transformations 
$\phi \rightarrow e^{i \alpha} \phi$ and $B \rightarrow B+ \beta$, with $\alpha, \beta$ being
different, independent constants. We will see further evidence of this global symmetry below when
we find a conserved current arising from the equations of motion of the $B$ field. In the usual 
spontaneous symmetry breaking process applied to a $U(1)$ global symmetry which is made local, 
the Goldstone boson of the original global $U(1)$ symmetry completely disappears -- it is ``eaten
by the massless gauge boson which then becomes massive. In the present case we will find that there is
some remnant of the Goldstone boson since the original global symmetry $U(1) \times U(1)$
was larger than the local $U(1)$ which was gauged.

Now, we briefly review some previous work dealing with scalar gauge fields or other non-traditional
ways of gauging a symmetry without (only) a vector gauge boson. The first example of 
a scalar field associated with the concept of gauging a symmetry is the paper by Stueckelberg \cite{Stueckelberg}
who investigated a compensating scalar field to construct gauge invariant mass terms. This work
used a restricted gauge transformations with the gauge parameter $\Lambda$ satisfying the free Klein Gordon 
equation for a certain mass. A modern review of the Stueckelberg approach is given in \cite{review}.
Next in \cite{guen1} gauge invariant mass terms were considered without any restriction on the 
gauge parameter $\Lambda$, and with the possibility of coupling to a current which was the gradient of a scalar 
field. The paper \cite{guen1} also studied the canonical formalism i.e. constraint structure.
In the work \cite{kato1} the idea of gauging the dual symmetry of E\&M \cite{jackson} was examined. It was found that
one could gauge the dual symmetry by introducing a scalar gauge field (see also \cite{saa} for an alternative
view of gauging the dual symmetry of E\&M). Reference \cite{chaves} gives a variant of the
Standard Model using vector and scalar gauge fields. In the work \cite{kato2} a modification of the usual gauge
procedure was proposed which allowed gauge fields of various ranks -- scalar (rank 0), vector (rank 1), and
higher tensors (rank 2 and higher). In \cite{guen2} scalar gauge fields were used to couple electromagnetism to 
a ``global charge" carried by a complex scalar field with only global phase invariance. Although the 
complex scalar fields considered in \cite{guen2} only had a global phase symmetry the 
gauge fields still had the full local gauge invariance. The use of scalar gauge fields 
also allows for the introduction of dynamical coupling constants 
in a self consistent fashion \cite{guen3}. Finally, scalar fields represented by a 
unitary matrix were used by Cornwall to formulate massive gauge theories 
in a gauge invariant fashion \cite{Cornwall}. 

We now fix, as far as possible, the character of the $c_i$'s in \eqref{u2}. First $c_1$, $c_2$ and $c_5$
must be real since $D^A _\mu \phi (D ^{A \mu} \phi ) ^*$, $D^B _\mu \phi (D ^{B \mu} \phi ) ^*$
and $(A_\mu - \partial _\mu B)(A^\mu - \partial ^\mu B)$ are real. Next $c_3$ and $c_4$ must be complex conjugates
(i.e. $c_3 = c_4 ^*$) in order that the combination of the two crossed covariant derivative terms in \eqref{u2} 
(i.e. the terms $D^A _\mu \phi (D ^{B \mu} \phi )^*$ and $D^B _\mu \phi (D ^{A \mu} \phi )^*$) be real.
Finally we require that $(c_1+c_2+c_3+c_4) = (c_1 +c_2 + Re[c_3 + c_4]) =1$. This condition ensures that the 
kinetic energy term for the scalar field $\phi$ has the standard form $\partial _\mu \phi \partial ^\mu \phi ^*$.
One could accomplish this as well by rescaling $\phi$, but here we chose to accomplish this by placing conditions 
on the $c_i$'s. Taking into account these conditions (and in particular writing out $c_3$ and $c_4$
in terms of their real and imaginary parts $c_3 =a+ib$ and $c_4 = a-ib$) the Lagrangian in \eqref{u2} becomes
\begin{eqnarray}
\label{u3}
{\cal L} &=& \partial _\mu \phi \partial ^\mu \phi ^* - V(\phi) - \frac{1}{4} F_{\mu \nu} F^{\mu \nu} 
+ c_5 A_\mu A^\mu + c_5 \partial _\mu B \partial ^\mu B -2 c_5 A_\mu \partial ^\mu B  \nonumber \\
&+& i e [\phi \partial _\mu \phi^* - \phi ^* \partial_\mu \phi] \left( (c_1 +a) A^\mu + (c_2 +a) \partial _\mu B \right) \\
&+& e^2 \phi \phi^* \left(  c_1 A_\mu  A^\mu + c_2 \partial _\mu B \partial ^\mu B + 2  a \partial _\mu B A^\mu \right) 
- e b \partial_\mu (\phi ^* \phi) (A ^\mu - \partial ^\mu B) ~.\nonumber
\end{eqnarray}
There are several interesting features of the Lagrangian in \eqref{u3}. First, the vector gauge field, $A_\mu$,
has a mass term (i.e. $c_5 A_\mu A^\mu$) which is allowed by the extended gauge symmetry \eqref{gauge-trans}.
Thus in addition to the vector gauge field developing a mass through the term
$e^2c_1 \phi \phi^* A_\mu A^\mu$ if $\phi$ develops a vacuum expectation value (i.e.
if $\langle \phi \phi ^* \rangle = \rho _0^2$ with $\rho _0$ a constant), there is
now an additional potential mass term for the vector gauge field, even in the absence of spontaneous
symmetry breaking via $\phi$. Second, the scalar gauge field appears to be a dynamical field
through the presence of two possible kinetic energy terms. The term $c_5 \partial _\mu B \partial ^\mu B$
is the standard kinetic energy term for a scalar field, especially if one chooses $c_5 = 1/2$. Also,
the term $c_2 e^2 \phi \phi^* \partial _\mu B \partial ^\mu B$ takes the form of a kinetic energy term
if $\phi$ develops a vacuum expectation value. Third, the term 
$- e b \partial_\mu (\phi ^* \phi) (A ^\mu - \partial ^\mu B)$ will lead to $C$ and $CP$ violation. 
This point will be discussed further in section \eqref{C-CP}. Fourth, 
unlike standard scalar QED, the strength of the seagull interaction of the vector gauge field
with the charged scalar matter field (i.e. the term $c_1 e^2 \phi \phi^*  A_\mu  A^\mu$) 
is independent of the strength of the one gauge particle emission/absorption 
interaction terms (i.e. the term
$i (c_1 +a) e [\phi \partial _\mu \phi^* - \phi ^* \partial_\mu \phi] A^\mu$ with the independence
coming from the presence of the constant $a$ which is not present in the seagull term). 

We now write down the Euler-Lagrange equations of motion for $A_\mu$, $B$ and $\phi$ in turn
and discuss some of the interesting features from this scalar gauge procedure. The Euler-Lagrange
equation for $A_\mu$ following from \eqref{u3} is
\begin{eqnarray}
\label{Au}
\partial _\nu F^{\mu \nu} &=& i e (c_1 +a) [\phi \partial ^\mu \phi^* -\phi^* \partial ^\mu \phi ]
-e b \partial ^\mu [\phi^* \phi] \nonumber \\
&+& 2[c_1 e^2 (\phi ^* \phi ) + c_5 ]A^\mu + 2[a e^2 (\phi^* \phi) - c_5] \partial ^\mu B ~.
\end{eqnarray}
The equation of motion for $B$ following from \eqref{u3} is
\begin{eqnarray}
\label{B}
\partial _\nu \left[ (2 c_5 + c_2 e^2 \phi^* \phi) \partial ^\nu B -2 (c_5 - a e^2 \phi^* \phi) A^ \nu \right] =
&-& i e(c_2 +a) \partial _\nu (\phi \partial ^\nu \phi ^* - \phi^* \partial ^\nu \phi ) \nonumber \\
&-& e b \partial _\nu \partial ^\nu (\phi ^* \phi)  ~.
\end{eqnarray}
Note that in \eqref{B} the term $2 c_5 \partial _\nu \partial ^\nu B = 2 c_5 \Box B$ looks like a standard
Klein Gordon type contribution for a real scalar field. Thus it appears that in this
gauge principle one generates not only a vector gauge field but also a real scalar field which are both fields 
arising from the gauging procedure/principle. Notice that in \eqref{B}  the phase of the complex scalar field appears 
through the term $\partial _\nu (\phi \partial ^\nu \phi ^* - \phi^* \partial ^\nu \phi )$. If we assume 
spontaneous symmetry breaking (SSB) (i.e. the complex scalar field develops a vacuum expectation value
like $\langle \phi^* \phi \rangle = \rho _0 ^2$) and if we choose to work in unitary gauge, such terms disappear. 
Also in the case of SSB where at low energies $\phi ^* \phi \approx \langle \phi^* \phi \rangle = \rho_0 ^2$ 
(i.e. $\phi ^* \phi$ is approximately constant), we see that the gauge field $B$ 
is a physical field due to the way it appears in the physical unitary gauge.
We will discuss this point further below in section \eqref{unitary-gauge}.

Notice that equations \eqref{Au} and \eqref{B} lead to two independent conservation laws of the form 
$\partial _\mu j^\mu =0$. Taking the divergence of \eqref{Au}, we get the conserved current $j_A ^\mu$
\begin{equation}
\label{A current}
j_A^\mu =  i e (c_1 +a) [\phi \partial ^\mu \phi^* -\phi^* \partial ^\mu \phi ]
-e b \partial_\mu [\phi^* \phi] + 
[2 c_1 e^2 (\phi ^* \phi ) +  2 c_5 ]A^\mu + 2[a e^2 (\phi^* \phi) - c_5] \partial ^\mu B ~.
\end{equation}
Equation \eqref{B} is already in the form of a conservation law of another current, the current 
$j_B^\mu $ given by,
\begin{equation}
\label{current of B}
j_B^\mu = i e (c_2 +a) [\phi \partial ^\mu \phi^* -\phi^* \partial ^\mu \phi ]
+e b \partial ^\mu [\phi^* \phi] +
[2 c_2 e^2 (\phi ^* \phi ) +  2 c_5 ]\partial^\mu B + 2[a e^2 (\phi^* \phi) - c_5] A ^\mu  ~.
\end{equation}
The conserved current \eqref{current of B} can be understood as the Noether current that 
arises from the global symmetry $B \rightarrow B + constant$. The current $j_B ^\mu$ is 
connected with one of the global $U(1)$'s from the original $U(1) \times U(1)$
global symmetry. Finally the equations of motion for $\phi$ are 
\begin{eqnarray}
\label{phi}
&&\partial _\mu \left[ \partial^\mu \phi +ie \phi (c_1 A^\mu + c_2 \partial ^\mu B) -e b \phi (A^\mu -\partial ^\mu B)
+i a e \phi (A^\mu + \partial ^\mu B) \right] =   \nonumber \\
&-& \frac{\partial V}{\partial \phi^*} + c_1 (e^2 A_\mu A^\mu \phi -i e \partial _\mu \phi A^\mu ) 
+ c_2 (e^2 \partial_\mu B \partial^\mu B \phi - i e \partial _\mu \phi \partial^\mu B ) \\
&-& e b \partial _\mu \phi (A^\mu -\partial ^\mu B) 
- i a e \partial _\mu \phi (A^\mu + \partial ^\mu B) +2 a e^2 \partial _\mu B A^\mu \phi  \nonumber~.
\end{eqnarray}
The equation of motion for $\phi ^*$ is the complex conjugate of \eqref{phi}.

\section{Particle content and the generalized unitary gauge}
\label{unitary-gauge}
We devote this section to the discussion of the physical reality of the newly introduced scalar gauge field $B(x)$ 
and to discuss the particle content of the theory when we have spontaneous symmetry breaking i.e. when the
scalar field $\phi$ develops a vacuum expectation value due to the form of the potential $V (\phi )$ in \eqref{u2} 
\eqref{u3}. At first glance one might conclude that $B(x)$ is not a physical field -- it appears that one could
``gauge" it away by taking $\Lambda = B(x)$ in \eqref{gauge-trans}. However one must be careful since this would
imply that the gauge transformation of the field $\phi$ would be of the form $\phi \rightarrow e^{i e B} \phi$ i.e.
the phase factor would be fixed by the gauge transformation of $B(x)$. In this 
situation one would no longer to able to use the usual unitary gauge transformation to eliminate 
the Goldstone boson in the case when one has spontaneous symmetry breaking.

The unitary gauge is the standard procedure to find the particle content of a spontaneously broken theory. 
Let us recall how the unitary gauge works: One writes the complex scalar field as an amplitude and phase -- 
$\phi(x) = \rho (x) e^{i \theta (x)}$. The two fields $\rho(x)$ and $\theta(x)$ represent the initial 
fields of the system. If $\phi (x)$ develops a VEV due to the form of the potential, $V (\phi )$, 
then one can transform to the unitary gauge  $\phi \rightarrow e^{ie \Lambda (x)} \phi(x)$ with $\Lambda =-\theta (x) /e$. 
In this way one removes the field $\theta (x)$ (which is ``eaten" by the gauge boson) 
and is left with only the $\rho (x)$ field. With the introduction of the 
scalar gauge field, $B(x)$, one no longer can gauge away {\it both} $\theta (x)$ and $B(x)$, and in the end one
is left with some real, physical field which is some combination of the original $B(x)$ and $\theta (x)$. Thus some
hint of the Goldstone boson survives in our extended gauge process as a physical field. This is 
supported by the fact that although the local symmetry of \eqref{u2} is $U(1)$ the original
global symmetry is $U(1) \times U(1)$. The evidence of this additional $U(1)$ can be seen in the
conserved current coming from the equation of motion of $B$ given in \eqref{current of B}. The SSB
of this $U(1) \times U(1)$ symmetry leads two Goldstone boson -- one which is ``eaten" by the
vector gauge boson thus giving the vector gauge boson a mass, and the other one remains (at least initially)
as a Goldstone boson.

Let us now define exactly what will be the generalization of the unitary gauge appropriate to 
the situation here. In the presence of spontaneous symmetry breaking and where the field
$\phi$ develops a VEV the unitary gauge eliminates cross terms like $A_\mu \partial ^\mu \theta$
from the Lagrangian (see for example \cite{Chen}). In the present case the cross terms 
between  the vector  field $A_\mu$ and the scalars (in our case $B$ and $\theta$) are more involved. 
Explicitly the relevant cross terms that we wish to eliminate by a generalized unitary gauge are 
\begin{equation}
\label{cross terms}
{\cal L}_{cross} =  -2 c_5 A_\mu \partial ^\mu B  
+ i e (c_1 +a) [\phi \partial _\mu \phi^* - \phi ^* \partial_\mu \phi] A^\mu
+ 2 a e^2 \partial _\mu B A^\mu \phi^* \phi ~.
\end{equation}
It is obvious why the first and third terms in the above equation are denoted as cross terms since they
have the form $A_\mu \partial ^\mu B$. To see why the second term above is considered a cross term between
$A_\mu$ and $\theta$ in the presence of SSB (i.e. the scalar field develops a VEV $\langle \phi \phi^* \rangle = \rho_0 ^2$ 
where $\rho _0$ is a constant) we begin by approximating the scalar field as $\phi (x) \approx \rho_0 e^{i \theta (x)}$.
With this the scalar current becomes 
$[\phi \partial _\mu \phi^* - \phi ^* \partial_\mu \phi] \approx 2 \rho _0 ^2 \partial_\mu \theta$. We
have used the assumption that the amplitude of the scalar field is approximately constant --
$\rho (x) \approx \rho _0$. Putting all this together show that the second term in 
\eqref{cross terms} is a cross term between $A_\mu$ and $\theta$ of the form $A_\mu \partial ^\mu \theta$. 
Thus \eqref{cross terms} becomes
\begin{equation}
\label{cross terms 2}
{\cal L}_{cross} = 2  A_\mu \partial ^\mu \left( - c_5 B  
+ e c_1 \rho_0 ^2  \theta +  a e \rho_0 ^2  \theta + a e^2  \rho_0 ^2  B \right) ~.
\end{equation}
It is this more complex cross term that we want to eliminate via some generalized 
unitary gauge. Defining $F (x) =-c_5 B  + c_1 e  \rho_0  ^2  \theta  + a e \rho_0 ^2  \theta + a e^2 \rho_0 ^2  B$,
one can see that the cross term in \eqref{cross terms 2} takes the form $\propto A_\mu \partial ^\mu F$
which is similar to the more common form  $\propto A_\mu \partial ^\mu \theta$ \cite{Chen}.   
By means of a gauge transformation ( i.e. $\theta \rightarrow \theta + e\Lambda$,
$B\rightarrow B-\Lambda$) we can take some initial non-zero value $F=F_0$, and always arrive at a
gauge $F=0$. From \eqref{cross terms 2} one can check this is possible by choosing 
the gauge function as $\Lambda = -F_0/(c_5 + c_1e^2 \rho_0 ^2)$. In this physical gauge, with $F=0$, we can 
solve the $\theta$ field in terms of the $B$ field as
\begin{equation}
\label{physical gauge}
\theta = \frac{c_5 - a e^2 \rho_0 ^2}{e \rho_0 ^2(c_1 + a)} B ~. 
\end{equation}
What \eqref{physical gauge} shows is that $\theta$ and $B$ are not independent fields -- one is fixed in
terms of the other. There is therefore only one physical scalar field in this generalized unitary gauge
which one can call either $\theta$ or $B$.  The above is different from the normal gauge procedure in the presence of
symmetry breaking where the $\theta (x)$ field completely disappears. Here there is some left over hint of the
Goldstone boson which we may call $B(x)$ (as we do here) or $\theta (x)$. At this stage the mixed $\theta/B$ field 
is massless and thus could be thought of as a true, massless Goldstone boson. However in section \eqref{B-potential}
we will show that it is possible to add to the Lagrangian from \eqref{u3}, non-derivative potential terms for
the $B$ field. These terms will include a mass term and power law interaction terms. 

After replacing $\theta$ in terms of $B$, from \eqref{u3} one sees that when $\phi$ develops an expectation value, 
the kinetic term for $B$ takes the form 
\begin{eqnarray}
\label{kinetic term B}
\left( c_5 +c_2  e^2  \rho _0 ^2 + 
2 \frac{(c_2  + a )(c_5 - a e^2 \rho_0  ^2 )}{\rho_0 ^2(c_1  + a )} \right) \partial_\mu B \partial^\mu B
\end{eqnarray}
To put \eqref{kinetic term B} in the canonically normalized form 
$\frac{1}{2}\partial_\mu \bar{B} \partial^\mu \bar{B}$, one should define $\bar{B}$ as
\begin{equation}
\label{bar-b}
\bar{B}= \left[ \sqrt{2 \left( c_2 \rho _0 ^2 e^2 + c_5 + 2 
\frac{(c_2  + a ) (c_5 - a e^2 \rho _0^2)}{\rho _0 ^2(c_1  + a )} \right)} \right] B  = f_B B ~, 
\end{equation}
where $f_B$, defined by the above equation. Depending on the value of these constants the normalized scalar gauge field,
$\bar{B} (x)$, may be larger or smaller compared to the original scalar gauge field, $B(x)$. At this point
after having imposed the generalized unitary gauge as defined above we have a remaining scalar gauge field
which we can denote either as $\theta$ or $B$ and which is massless. Neither $\theta$ nor $B$ has a mass
term if one looks at the Lagrangian \eqref{u3} after the imposition of the generalized unitary gauge.
Thus one can say that in this extended gauge procedure there is a remaining Goldstone boson like particle.
Below, in section \eqref{B-potential} we will show that one can add non-derivative, potential terms to the
Lagrangian in \eqref{u3} which give mass and non-derivative self interaction terms to the $B$ field so that the 
$B$ field can be viewed as a massive remnant of the usual Goldstone boson.

The above discussion has been carried out under the assumption that the scalar field, $\phi$, undergoes SSB and develops a
non-zero VEV, $\rho _0 \ne 0$. One can ask what happens when there is no SSB and $\rho _0 =0$? In this case one can see
that \eqref{cross terms 2} reduces to just ${\cal L}_{cross} = -2 c_5 A_\mu \partial ^\mu B$. There is now no
connection between $B$ and $\theta$ such as that given in \eqref{physical gauge}.
It thus appears that in order to make the cross term vanish one simply needs to use the gauge
transformation $B \rightarrow B-\Lambda$ with $\Lambda (x) = B(x)$ to transform $B(x)$ away. 
This leads to an apparent puzzle in regard to the counting of degrees of freedom between the
case without SSB and the case with SSB. In the case {\it with} SSB there are
five dergees of freedom: three degrees of freedom from the massive vector field $A_\mu (x)$, one degree of 
freedom from $\rho (x)$ and one degree of freedom from the mixture of $\theta (x) / B(x)$. 
In the case {\it without} SSB where $\rho _0 =0$ it at first {\it appears} that one has 
only four degrees of freedom: two degrees of freedom from the {\it apparently} massless vector field, $A_\mu (x)$, 
and two degrees of freedom from the complex scalar field, $\phi (x)$ which we have denoted $\rho (x)$ 
and $\theta (x)$. We have used the gauge freedom to gauge away the $B (x)$. However in this case
{\it without} SSB one has already used the gauge symmetry to gauge away $B(x)$ and thus one can no longer
use this to reduce the degrees of freedom for the $A_\mu (x)$ down to two. Note also that in \eqref{u2}
or \eqref{u3} there are terms like $c_5 A_\mu A^\mu$ which are allowed by the extended gauge procedure and 
such terms do represent mass terms for $A_\mu$.

\section{Breaking of $C$ and $CP$ symmetries}
\label{C-CP}
In this section we study the violation of $C$ and $CP$ that occurs in the scalar gauge field
model with the Lagrangian given in \eqref{u2}. Under $C$ symmetry the scalar field, $\phi$, is replaced 
by its complex conjugate, and both vector and scalar gauge fields change sign,
\begin{equation}
\label{C}
\phi \rightarrow  \phi^* ~~~;~~~ A_\mu \rightarrow -A_\mu  ~~~;~~~
B \rightarrow - B ~.
\end{equation} 
As we have seen  $c_3$ and $c_4$ must be complex conjugates
(i.e. $c_3 = c_4 ^*$). However if they are complex (i.e. $c_3 = a + i b$ , $c_4 ^* = a -i b$) then the 
charge conjugation symmetry of \eqref{C} is broken in the Lagrangian \eqref{u2} and in the equations of motion 
by $b$ the imaginary part of $c_3$ and $c_4$. For example one can see that under the charge conjugation 
symmetry the part of the current proportional to $b$ appearing in \eqref{u3} namely 
\begin{equation}
\label{c-violate}
e b \partial_\mu (\phi ^* \phi) (A ^\mu - \partial ^\mu B) ~,
\end{equation}
transforms differently under charge conjugation in \eqref{C} than the other 
pieces. As a consequence the currents by a particle and its associated antiparticle will not be 
exactly opposite to each other if the gradient term, $\partial_\mu (\phi ^* \phi)$, is non-vanishing 
i.e. for space time dependent $\phi \phi^*$. 

Turning to the transformation of the fields, $\phi$, $B$, $A_\mu$, under parity (i.e.
$ t \rightarrow t$ and ${\bf x} \rightarrow - {\bf x}$) we see that the fields 
$\phi$ and $B$ transform as scalars under parity and $A_\mu$ transforms as a normal vector 
under parity. Mathematically this amounts to ($i=1,2,3$)
\begin{eqnarray}
\label{Parity}
 A_0(x^i, t) \rightarrow A_0(-x^i, t)  ~~~;~~~ A_i(x^i, t) \rightarrow -A_i(-x^i, t) \nonumber \\ 
 \phi(x^i, t) \rightarrow  \phi(-x^i, t)~~~;~~~ B(x^i, t) \rightarrow  B(-x^i, t)  ~.
\end{eqnarray}
Thus the action following from the Lagrangian in \eqref{u2} respects parity. Combining this
result with the violation of charge conjugation symmetry coming from the term \eqref{c-violate} one
finds that the Lagrangian violates $CP$ as well as $C$. 

\section{Non-derivative interactions for the $B$ field}
\label{B-potential}

In this section we show that it is possible to add $B$-dependent terms to the scalar field
potential $V(\phi )$. The simplest example of such an addition leads to the sine-Gordon
field equation for $B$ in the unitary gauge. This implies that the $B$ field has soliton/kink like solutions.
We also show that in the case when $\phi$ undergoes symmetry breaking and develops a vacuum expectation 
value (VEV) that the interaction strength of the $B$ field is suppressed by this VEV. Thus for a large VEV 
one has a natural explanation for why the $B$ field should be weakly interacting.  

The Lagrangian in \eqref{u2} has a polynomial self interaction potential, $V(\phi)$, with terms
of the form $m^2 \phi^* \phi$ and $\lambda (\phi^* \phi )^2$, but the interactions of 
$B$ in \eqref{u2} involve only $\partial _\mu B$. One can introduce non-derivative interaction terms 
for $B$ by noting that terms like $e^{i n e B}(\phi)^n$ and $e^{- i n e B}(\phi ^*)^n$ are invariant under the
gauge transformation of $\phi$ and $B$ given in \eqref{gauge-trans}. If one wants to restrict 
attention to renormalizable theories, then one might limt such terms to be no more than quartic 
in $\phi$ or $\phi^*$. It is not clear that such a theory would be renormalizable since the scalar
gauge field $B$ appears in an exponential and thus involves all power of $B$, but by having
$n \le 4$ means that at least in regard to simple power count of the field $\phi$ the theory is renormalizable. 
In any case taking $n \le 4$ one can write down the following general interacting potential
for $\phi$ and $B$  
\begin{eqnarray}
\label{B dependent potential}
V( e^{i eB} \phi) = - m^2 \phi \phi^* + \lambda (\phi \phi^*)^2 + \lambda_1 e^{i eB} \phi + \lambda^*_1 e^{-i eB} \phi^*
+ \lambda_2 e^{i 2eB} \phi^2 +  \lambda_2^* e^{-i 2eB} (\phi^*)^2 \nonumber \\
+ \lambda_3 e^{i 3eB} \phi^3 + \lambda_3^* e^{-i 3eB} (\phi^*)^3 + \lambda_4 e^{i 4eB} \phi^4 + \lambda_4^* e^{-i 4eB} 
(\phi^*)^4  ~.
\end{eqnarray}
If the $\lambda_i$ 's, have an imaginary part, then those terms will violate charge conjugation symmetry. Note also 
that one could generalize \eqref{B dependent potential} further by terms of the form
$e^{ie B} (\phi)^2 (\phi^*)$ or $e^{-ie B} (\phi )(\phi^*)^2$. 

We now specialize \eqref{B dependent potential} to terms that are only linear in $\phi$ or $\phi^*$ by
taking $\lambda _i =0$ for $i \ge 2$ so that we have 
\begin{equation}
\label{B dependent potential-2}
V( e^{i eB} \phi) = - m^2 \phi \phi^* + \lambda (\phi \phi^*)^2 + \lambda_1 e^{i eB} \phi + \lambda^*_1 e^{-i eB} \phi^*
\end{equation}
We write the scalar field in the polar form $\phi (x) = \rho (x) e^{i \theta (x)}$ $\phi$ and take the gauge 
to be that defined by \eqref{physical gauge} so that $\theta(x)$ is given in terms of $B$.
Finally, we re-express everything in terms of the canonically normalized field $\bar{B}$ defined before, 
which implies that $\theta+eB= K\bar{B}$ with $K$ being defined as
\begin{equation}
\label{K}
K  =\frac{c_5 + \rho _0 ^2 e^2c_1}{f_B \rho_0 ^2(c_1 e + ae)} ~,
\end{equation}
and with $f_B$ being defined above via \eqref{bar-b}.
With these choices the potential in  \eqref{B dependent potential} becomes
\begin{equation}
\label {physical gauge potential}
V( \bar{B} , \rho) = - m^2 \rho ^2 + \lambda \rho ^4 + (\lambda_1 e^{i K\bar{B}} \ + \lambda^*_1 e^{-i K\bar{B}}) \rho ~.
\end{equation}
The first two terms in \eqref{physical gauge potential} are the standard scalar symmetry breaking 
potential which give rise to a vacuum expectation value of $|\phi | \approx \rho _0 = \sqrt{m ^2 /2 \lambda}$.
Now if we assume that the last two terms involving $B$ do not shift the value of this vacuum expectation value
to any great degree, and writing $\lambda _1 = \alpha _1 e^{i \omega _1}$ we see that near the vacuum
value $\phi \approx \sqrt{m ^2 /2 \lambda}$ the potential in \eqref{physical gauge potential} becomes
\begin{equation}
\label {sine Gordon}
V( \bar{B} ) = -\frac{m^4}{4 \lambda} + 2\alpha_1  \rho _0 \cos (K\bar{B}+\omega_1) ~.
\end{equation}  
This is of the form of the sine-Gordon equation, which has the interesting feature of having 
topological kink/soliton solutions \cite{SG}. If we expand the term $2\alpha_1  \rho _0 \cos (K\bar{B}+\omega_1)$
in \eqref{sine Gordon} in ${\bar B}$ we find terms proportional to ${\bar B} ^2$, ${\bar B} ^4$ 
${\bar B} ^{2n}$ ... These represent a mass term, quartic self interaction term and higher polynomial
interaction terms. Thus the field ${\bar B}$ or $B$ (or more precisely the mixture of $B (x)$ and $\theta (x)$
are not true Goldstone bosons since by adding potential terms like \eqref{physical gauge potential}
or \eqref{sine Gordon} one can give mass terms to these fields, where in contrast a true Goldstone
boson is massless. 

We next turn to the question of the interaction strength of $B$. From looking at the covariant derivative
$D^B _\mu$ from \eqref{cov-ab} one would think that the coupling strength should be determined by $e$. 
However with symmetry breaking, where the scalar field develops the VEV, 
$|\phi | \approx \rho _0 = \sqrt{m ^2 /2 \lambda}$, this expectation is changed. 
The relevant field to consider is not $B$, but the canonically normalized field $\bar{B}$.
The strength of the interaction of $\bar{B}$ with the scalar $\phi$ is not $e$ (as is the case 
with $B(x)$) but rather $e/f_B$. If $f_B$ is big (as is the case if the VEV, $\rho _0$, and/or
$c_2 , c_5$ are large) the coupling of the $B$ particles will be reduced. 
This is reminiscent of a similar mechanism occurring the axion \cite{axion}. 

\section{Discussion and Conclusions}

In this article we have presented a gauging procedure using a scalar gauge field as well as
the more standard vector gauge field -- our gauge fields were $B(x)$ and $A_\mu (x)$. In this
paper the symmetry we gauged was the Abelian $U(1)$ symmetry of a scalar matter field $\phi (x)$.
However the gauging process presented here can be extended to the case where one starts with
a spinor field, $\Psi (x)$, rather than a scalar matter field, $\phi (x)$. It is also possible
to extend the gauge procedure presented here to the case of non-Abelian symmetries. In the case
of non-Abelian symmetries one that the procedure presented here has some close similarities 
to the work of Cornwall in \cite{Cornwall}. We will return to this point of a non-Abelian
version of the present gauge procedure in a future paper.

In this paper we have focused mainly of simply laying out our alternative gauge 
procedure with both scalar and vector gauge fields, without stressing too much any
physical applications. However, there are hints of interesting possible physical 
applications for this gauge procedure and the scalar gauge field, $B(x)$. First we mention
the natural emergence of $C$ and $CP$ violation in this model arising from the terms
in the Lagrangian \eqref{u3} of the form $e b \partial_\mu (\phi ^* \phi) (A ^\mu - \partial ^\mu B)$.
Such a new source of $CP$ violation could be applied to baryon-genesis in the early
Universe. Further, in the model presented here the $CP$ violation can be correlated to the time 
dependence of a scalar field, which fits very much the modern approaches to cosmology of the 
early universe, in particular the Higgs inflation scenarios \cite{Higgs inflation}. 
Second, one can get interesting but non-standard coupling between the matter fields ($\phi (x)$ in
this paper) and the gauge fields ($B(x)$ and $A_\mu (x)$). For example by selecting 
the constants in the Lagrangian \eqref{u3} such that $c_1=c_2 = -a$ one can get rid of the
conventional current terms ($[\phi \partial ^\mu \phi^* -\phi^* \partial ^\mu \phi ]$) 
and is left with a current of the form $\partial_\mu [\phi^* \phi]$ which represents the ``maximal $C$ 
and $CP$ violating case".  Also for different choices for the various constants $c_1, c_2, a, b, c_5$
there are terms, such that the strength of the seagull interaction of the vector gauge field with the charged 
scalar matter field, which are of independent to the strength of the one gauge particle emission/absorption 
interaction terms. Third, one can introduce non-derivative couplings of the scalar gauge field
$B(x)$ using the gauge invariance of $e^{i e B} \phi$ to arrive at potentials of the form 
given in \eqref{B dependent potential}. By assuming a a symmetry breaking potential for the scalar field,
$\phi (x)$, it is possible to get a sine-Gordon like equation \eqref{sine Gordon} for the scalar gauge field $B(x)$
if $\phi (x)$ remains close to its vacuum expectation value $\phi \approx \sqrt{m ^2 /2 \lambda}$.
One final potential application of the scalar gauge field $B(x)$ might be as a potential explanation
of the proton spin/angular momentum puzzle \cite{lorce} i.e. the fact that very little of the spin of
the proton comes from the spin of the valence quarks. Thus the spin of the proton (or nucleon in general) 
must reside in some combination of the gauge field \cite{sing-1998} and/or orbital angular momentum
\cite{sing-2000}. Extending the gauge field structure, as in this paper, might provide an additional
potential source for the spin of the proton/nucleon.


\begin{thebibliography}{99}

\bibitem{Stueckelberg} E.C.G. Stueckelberg, Helv. Phys. Acta 11, 225 (1938)

\bibitem{review} H. Ruegg and M. Ruiz Altaba, e-Print: arXiv:0304245[hep-th]

\bibitem{guen1} E. Guendelman, Phys. Rev. Lett. {\bf 43}, 543-545 (1979)

\bibitem{kato1} A. Kato and D. Singleton, Int. J. Theor. Phys. {\bf 41}, 1563-1572 (2002) 

\bibitem{jackson} J.D. Jackson, {\it Classical Electrodynamics}, 3$^{rd}$ edition section 6.11
(John Wiley \& Sons Inc. 1999)

\bibitem{saa} A. Saa, Class. Quant. Grav. {\bf 28}, 127002 (2011) 

\bibitem{chaves} M. Chaves and H. Morales, Mod. Phys. Lett. {\bf A13}, 2021 (1998);
M. Chaves and H. Morales, Mod. Phys. Lett. {\bf A15}, 197 (2000).

\bibitem{kato2} D. Singleton, A. Kato, and A. Yoshida. Phys. Lett. {\bf A330}, 326-337 (2004) 

\bibitem{guen2} E. Guendelman, Int. J. Mod. Phys. {\bf A28}, 1350169 (2013) 

\bibitem{guen3} E.I. Guendelman, R. Steiner, ``Confining Boundary conditions from dynamical Coupling Constants" 
e-Print: arXiv:1311.2536 [hep-th]

\bibitem{Cornwall} J. M. Cornwall, Phys. Rev. D 26, 1453 (1982).

\bibitem{Chen} Ta-Pei Cheng and Ling-Fong Li, {\it Gauge Theory of elementary Particle Physics}, 
Oxford Science Publications, Clarendon Press, Oxford, Oxford University Press, NY, 1988.

\bibitem{Higgs inflation} F. L. Bezrukov and M. Shaposhnikov, Phys. Lett. {\bf B659}. 703 (2008);
F. Bezrukov, Class. Quant. Grav. {\bf 30}, 214001 (2013).

\bibitem{SG} R. Rajaraman, {\it Solitons and Instantons: An Introduction to Solitons and 
Instantons in Quantum Field Theory}, pp. 34 –- 45 (North-Holland 1989)

\bibitem{axion} J. E. Kim and G. Carosi,  Rev. Mod. Phys. {\bf 82}, 557 (2010) 

\bibitem{lorce} E. Leader and C. Lorc{\'e}, 
``The angular momentum controversy: What's it all about and does it matter?" to be published Phys. Rept.;
arXiv:1309.4235 [hep-ph].

\bibitem{sing-1998} D. Singleton, Phys. Lett. {\bf B427}, 155 (1998).

\bibitem{sing-2000} D. Singleton and V. Dzhunushaliev, Found. Phys. {\bf 30}, 1093 (2000). 



\end{thebibliography}
\end{document}